\documentclass[preprint,cite,bm]{epl}

\usepackage{graphicx}
\usepackage{amsmath,amssymb,amsfonts}

\title{Fully self-consistent {\it GW} calculations for atoms and molecules}
\shorttitle{Fully SC-{\it GW} calculations for atoms and molecules} 
\author{Adrian Stan, Nils Erik Dahlen and Robert van Leeuwen}
\shortauthor{A. Stan \etal}

\institute{                    
  \inst{1} Rijksuniversiteit Groningen, Materials Science Centre, Theoretical Chemistry, Nijenborgh 4, 9747AG Groningen, The Netherlands.\\
}

\begin{document}

\newcommand{\be}{\begin{equation}}
\newcommand{\ee}{\end{equation}}
\newcommand{\bea}{\begin{eqnarray}}
\newcommand{\eea}{\end{eqnarray}}
\newcommand{\br}{\mathbf{r}}
\newcommand{\bx}{\mathbf{x}}
\newcommand{\GHF}{G_{\rm{HF}}}

\maketitle

\date{\today}

\begin{abstract}
We solve the Dyson equation for atoms and diatomic molecules
within the $GW$ approximation, in order to elucidate the effects of
self-consistency on the total energies and ionization potentials.
We find $GW$ to produce accurate energy differences although the 
self-consistent total energies differ significantly from the exact 
values. 
Total energies obtained from the Luttinger-Ward functional 
$E_{\rm{LW}}[G]$ with simple, approximate Green functions as input, are shown 
to be in excellent agreement with the self-consistent results. 
This demonstrates that the Luttinger-Ward functional is a reliable method for 
testing the merits of different self-energy approximations without the need to 
solve the Dyson equation self-consistently. Self-consistent $GW$ ionization 
potentials are calculated from the Extended Koopmans Theorem, and shown to be 
in good agreement with the experimental results.
We also find the self-consistent ionization potentials to be often better 
than the non-self-consistent $G_0W_0$ values.
We conclude that $GW$ calculations should be done self-consistently in
order to obtain physically meaningful and unambiguous energy
differences.
\end{abstract}

\section{Introduction} Green function methods have been used with great success to calculate a 
wide variety of properties of
electronic systems, ranging from atoms and molecules to solids. 
One of the most successful and widespread methods has been the
$GW$ approximation ($GW$A) \cite{hedin65}, which has produced excellent results
for band gaps and spectral properties of solids 
\cite{aryasetiawan98, godby98}, but
so far has not been explored much for atoms and molecules, although it has
been known that for atoms the core-valence interactions are described much more
accurately by $GW$ than Hartree-Fock (HF) \cite{shirley93}. Moreover,
the  $GW$ calculations are rarely carried out 
in a self-consistent manner, and the effect of self-consistency is for this 
reason still a topic of considerable debate \cite{ku02,delaney04}. In this 
paper we present self-consistent all-electron $GW$ (SC-$GW$) calculations for 
atoms and diatomic molecules. The reason for doing these calculations is
two-fold: Firstly we want to study the importance of self-consistency within 
the $GW$ scheme. Such calculations are usually avoided due to
the rather large computational effort involved. It has been suggested that 
self-consistency will in fact worsen the spectral properties, though 
calculations on silicon and germanium crystals indicate that this is not 
always the case \cite{ku02}. The second reason is that we aim to study 
transport
through large molecules and molecular chains, where it is essential to account
for the screening of the long range of the Coulomb interaction. The 
calculations on diatomic molecules are the first step in this direction.

The $GW$A is obtained by
replacing the bare Coulomb interaction $v$ in the exchange self-energy with the
dynamically screened interaction $W$, such that $\Sigma=-GW$. The screened
interaction also depends on the Green function, and one thus needs to solve
a set of coupled equations for $G$ and $W$. One usually goes
through only a single iteration of this scheme. 
With an initial Green function $G_0$ calculated from, e.g., the local density 
approximation (LDA), one calculates $W$ and $\Sigma$, and subsequently obtains
a new Green function from the Dyson equation. This scheme, known as 
the $G_0 W_0$ approximation, has produced good results for a wide
variety of systems \cite{aryasetiawan98}, but 
suffers from a dependence on the choice of the initial 
$G_0$. Moreover, observables like the 
total energy are not unambiguously defined, and can be calculated in several 
different ways.  These problems can be cured by performing 
self-consistent calculations \cite{baym62}, since the $GW$A 
is a $\Phi$-derivable approximation (see Fig.~\ref{fig1:gwaphigw}).
The fact that self-consistency removes these ambiguities does not imply that
the results are necessarily closer to the exact values. For the electron gas 
it was shown that self-consistency actually worsens the spectral properties, while the total
energy is in excellent agreement with Monte-Carlo results \cite{holm98}.
On the other hand, for a system of very localized interactions, 
SC-$GW$ produced poor results for
both total energies and spectral properties \cite{schindlmayr98}.
Furthermore, Delaney \textit{et al.} \cite{delaney04} recently published 
SC-$GW$ results for the ionization potential of the Be atom that were worse 
than those of $G_0 W_0$.
Calculations on the Si and Ge crystals have, however, shown that 
self-consistency leads to improved band gaps \cite{ku02}. 

\section{General formulation} In this paper, we study the importance of self-consistency in $GW$ for atoms 
and diatomic molecules.
We compare the self-consistent total energies to those obtained from the
Luttinger-Ward (LW) functional \cite{lw60} which was earlier
used to estimate the $GW$ total energy for atoms \cite{dahlen04b} and the 
electron gas \cite{almbladh99}.
The LW functional $E_{\rm{LW}}[G]$ is a variational energy functional
in the sense that 
$\delta E_{{\rm {LW}}}[G]/ \delta G = 0$, when $G$ is a self-consistent 
solution of the Dyson equation. This variational property suggests that 
evaluating $E_{\rm{LW}}$ on an approximate Green function obtained from, 
e.g., HF or LDA calculations will give a result very close to
the self-consistent value. This was earlier shown to be the case for the
second-order self-energy \cite{dahlen05b}, and investigating the stability of
the LW functional also for the 
$GW$A is an important goal of this paper. The previously published LW 
calculations \cite{dahlen04b} indicated that the $GW$ total energies are not 
very accurate, but the essential question is 
rather whether total energy differences are produced accurately. We have for 
this reason also calculated the binding curve of the H$_2$ molecule and
two-electron removal energies  $\Delta E = E_{N-2} - E_N$.

We use the finite temperature formalism, with a 
temperature $T$ (we are only considering the limit 
$T \rightarrow 0$) and a chemical potential $\mu$. 
The Green function depends on the imaginary time coordinate $\tau$, in the 
range $-\beta\leq \tau \leq \beta \equiv 1/k_B T$, where $k_B$ is 
the Boltzmann constant. It satisfies the Dyson equation
\begin{eqnarray}
&&\Bigg[ - \partial_{\tau} + \frac{\nabla^2}{2} - w({\bf r}) - v_H({\bf r}) + \mu \Bigg] G({\bf x}, {\bf x'}; \tau)= \nonumber \\
&&= \delta ( \tau ) \delta ({\bf x} - {\bf x'}) + \int_{0}^{\beta} \upd \tau_1 \int \upd {\bf x}_1 \Sigma[G] ({\bf x}, {\bf x}_1; \tau-\tau_1)\times G({\bf x}_1, {\bf x'}; \tau_1),
 \label{eq:Dyson} 
\end{eqnarray}
where $\bx=(\br,\sigma)$ denotes the space- and spin coordinates, $w(\br)$ is 
the external potential, 
$\Sigma[G]({\bf x}, {\bf x'}; \tau)$ is the self-energy and 
$v_H({\bf r})$ is the Hartree potential. The last two objects are functionals 
of the Green function, and the Dyson equation should therefore be solved 
self-consistently, together with the boundary conditions $G(\bx,\bx',
\tau-\beta)=-G(\bx,\bx';\tau)$  and $G(\bx,\bx';0^+)-G(\bx,\bx';0^-)= 
-\delta(\bx-\bx')$.
\begin{center}
\begin{figure}
\onefigure[width=9cm]{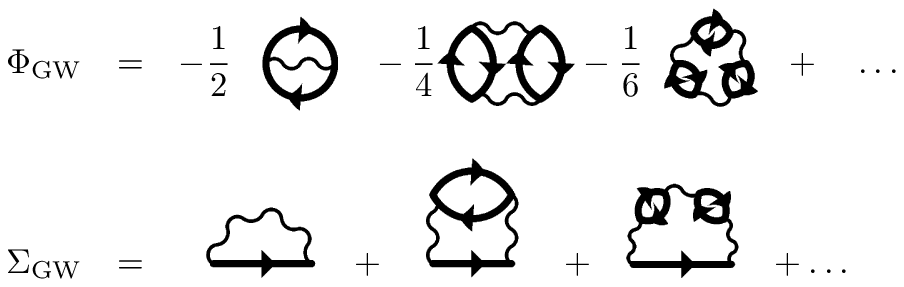} 
\caption{\label{fig1:gwaphigw} The $GW$ self-energy $\Sigma$ is the functional 
derivative of a functional $\Phi[G]$.}
\end{figure}
\end{center}
In the $GW$A (Fig.~\ref{fig1:gwaphigw}) the electronic self-energy is given by 
$\Sigma=-GW$ using the screened interaction $W=v+vPW$, where $v$ is the bare Coulomb interaction $1/|\br-\br'|$ and $P=GG$ is the polarizability 
\cite{hedin65}. The Green function is transformed into a $\tau$-dependent 
matrix by expanding
it in a basis of molecular orbitals obtained from an initial HF 
calculation. These molecular orbitals are linear 
combinations of Slater functions located on the atomic centers 
\cite{basisset}. 
The Green function, the $\Sigma[G]$ and the $W$  are peaked around the 
endpoints ($\tau=0$ and $\tau=\pm\beta$)  \cite{dahlen05b, ku02} so their
representation on an even-spaced grid is inconvenient. Instead, we used a
mesh which is dense around the end points \cite{ku02}. 

Since we calculate the Green function on the imaginary time axis, 
it is inconvenient to calculate the ionization potentials by finding the poles 
of the Green function in frequency space, $G(\omega)$. We have instead used the
extended Koopmans theorem (EKT) 
\cite{smith75} where the ionization potentials are found from the eigenvalue 
equation 
\begin{equation}
\sum_{ij} \Delta_{ij} u_{j}^{m}=-\lambda_{m} \sum_{j}\rho_{ij}u_{j}^{m},
 \label{eq:EKT}
\end{equation}
where $\Delta_{ij}=-\partial_\tau G_{ij}(\tau)|_{\tau=0}$, the density
matrix is given by $\rho_{ij}=G_{ij}(0^-)$ and the matrix indices refer to
the molecular orbital basis \cite{dahlen05b}. The eigenvalues $\lambda_m$
are interpreted as $\lambda_m = E_m^{N-1}-E_0^N+\mu$, i.e. the ionization 
potentials plus the chemical potential. The EKT is known to be exact for the 
lowest ionization energies, if the exact $\Delta$ and $\rho$ matrices are given
\cite{katrieldavidson}. For the HF approximation, the EKT eigenvalues obviously agree with
the poles of the HF Green function, and it is an unproven conjecture that 
these two methods will give the same value for the first ionization potential 
when the Green function is calculated self-consistently within a conserving
approximation.
The EKT has recently been used to calculate ionization
potentials for atoms and molecules from a self-consistent Green function using
the second order diagrams \cite{dahlen05b}. 

To calculate the SC-$GW$ total energy $E=T+V_{ne}+U_0+U_{\rm{xc}}$, 
we use the fact that the exchange-correlation part of
the interaction energy is given by
\be
U_{\rm{xc}}=\frac{1}{2} \sum_{ij} \int_0^\beta \upd \tau  \Sigma_{ij}(-\tau) 
G_{ji}(\tau), \label{eq:uxc}
\ee
and the kinetic energy $T$, nuclear-electron attraction energy $V_{ne}$ and
Hartree energy $U_0$ are trivially obtained from the density matrix $\rho$.
There are many other ways to calculate the total energy from a given Green
function, but only for a self-consistent solution of the Dyson equation
will these methods give the same result \cite{baym62}. 
One alternative is to calculate the  energy from variational functionals of the Green 
function. LW have shown \cite{lw60} that the total energy can be written as
\be
E_{\rm{LW}}[G] = \Phi[G] - U_0 - 
{\rm{Tr}} \big\{\Sigma G\big\}  
- \rm{Tr} \ln  [\Sigma - G_{H}^{-1} ]+\mu N
\ee
where $G_H$ is the Hartree Green function, and $\Sigma=\delta \Phi/\delta G$. 
The trace indicates an integration over the spatial coordinates and
 $\tau$ \cite{dahlen04b}, see also Eq.~(\ref{eq:uxc}). It is easily verified that 
$\delta E_{\rm{LW}}/\delta G=0$ when $G$ is a self-consistent solution
of the Dyson equation (\ref{eq:Dyson}). 
Hence, if we evaluate the LW functional on a simple input Green function,
we obtain a result close to the self-consistent energy, since we make an
error only to second-order in the deviation from the self-consistent $G$. 
This means that we have a computationally cheap way of 
obtaining self-consistent total energies. 
\begin{table}
\caption{Total energies (in Hartrees) calculated from SC-$GW$ compared to CI values and results from the LW functional and Galitskii-Migdal formula evaluated on $G_{\rm{HF}}$. }
\label{tab:totale}
\begin{largetabular}{lcccc}
\hline
System & $E_{\rm {SC}}^{GW}$ & $E^{GW}_{\rm{LW}}[G_{\rm {HF}}]$ &  
$E_{\rm{GM}} [G_{\rm {HF}}]$ & CI \\  
\hline
\chem{He}        &   -2.9278 &   -2.9277 & -2.9354 & -2.9037$^1$ \\
\chem{Be}        &  -14.7024 &  -14.7017 & -14.7405 & -14.6674$^1$ \\
\chem{Be^{2+}} & -13.6885 & -13.6885 & -13.6929 & -13.6556$^1$\\
\chem{Ne}        & -129.0499 & -129.0492 & -129.0885 & -128.9376$^1$\\
\chem{Mg}        & -200.1762 & -200.1752 & -200.2924 & -200.053$^1$\\
\chem{Mg^{2+}} & -199.3457 & -199.3453 & -199.3785 & -199.2204$^1$\\
\chem{H_2}     &  -1.1887 &  -1.1888 & -1.1985  & -1.133$^2$\\
\chem{LiH}       &   -8.0995 &   -8.0997 & -8.1113 & -8.040$^3$\\
\hline
\end{largetabular}
$^1$From Ref. \cite{chakravorty93}. $^2$From Ref. \cite{vanleeuwen94}.
$^3$From Ref. \cite{li03}.\\
\end{table}
The quality of the energies will 
ultimately be determined by the chosen self-energy approximation. 

Within a molecular orbital basis, the Dyson 
equation (\ref{eq:Dyson}) becomes a matrix equation. We 
introduce a reference Green function $G_0$ in order to write the equation on
integral form,
\bea
G_{ij}(\tau)&=& \delta_{ij} G_{0,i}(\tau) +\int_0^\beta \upd\tau_1\int_0^\beta
\upd\tau_2 \sum_k G_{0,i}(\tau-\tau_1)  \tilde \Sigma_{ik}(\tau_1-\tau_2) G_{kj}(\tau_2),
\label{eq:intdyson}
\eea
where $\tilde \Sigma = \Sigma[G]-\Sigma_0$, and $\Sigma_0$ is the self-energy
corresponding to $G_0$ \cite{dahlen05b}. We take $G_0$ and $\Sigma_0$ to be
the HF Green function and self-energy, but this choice is arbitrary. Using, 
e.g., LDA instead would not change any of the results.
The inverse temperature is chosen to have 
a sufficiently large value, typically larger than 100 a.u.. The value of the 
chemical potential is somewhat
arbitrary, but should be in the gap between the highest occupied
and the lowest unoccupied orbital. We checked that
the observables calculated from the resulting Green function did not depend on 
the choice of $\beta$ and $\mu$. 
The calculations on the molecules were done
at the experimental bond lengths. 

\section{Results} In Table \ref{tab:totale} we show the SC-$GW$ total energies 
of some atoms and small molecules. We have also included the $E_{\text{LW}}[\GHF]$
results, which are in spectacular 
agreement with the SC-$GW$ values. This
agreement is independent of the chosen basis set, and was earlier observed
also for the second-order diagrams \cite{dahlen05b}. The third column
shows the total energy calculated from $\GHF$
using the Galitskii-Migdal \cite{gm58} formula. In contrast to the LW results,
these are not in good agreement with the self-consistent energies. This clearly
demonstrates that different total energy functionals will not produce the same 
results when evaluated on a non-selfconsistent Green function (in this case, 
$\GHF$), and it also demonstrates the importance of using the variational
functionals for obtaining a result in agreement with the self-consistent 
values. 

\begin{figure}
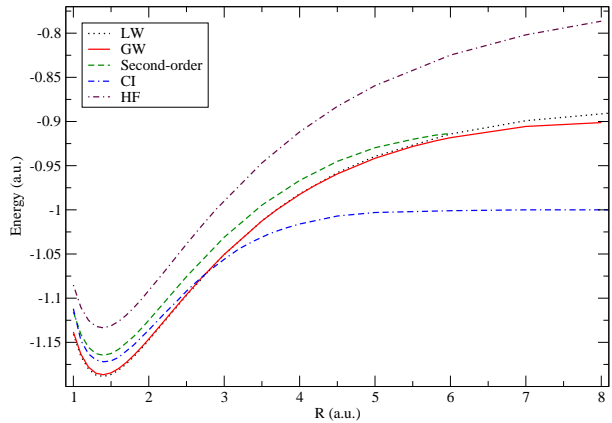

\onefigure[width=8cm]{fig2.eps} 
\caption{\label{fig:bindingh2}The total energy of the ${\rm H_2}$ molecule, 
as function of the interatomic distance, calculated within the second order, 
the self-consistent $GW$A, the $E_{\rm{LW}}^{GW}[\GHF]$ functional and
CI (from Ref.~\cite{vanleeuwen94}). For comparison, the HF results are also presented. }
\end{figure}

As a further test of the total energy functionals, we have calculated the 
total energy of the \chem{H_2} molecule for a range
of internuclear separations. Figure \ref{fig:bindingh2} shows the 
SC-$GW$ results together with the $E_{\rm{LW}}[\GHF]$ energy.
The curves agree closely up to $R\approx 5$ a.u., and the deviation remains
small even at $R=8$. The gradual increase in the deviation is due to the fact 
that the input $\GHF$ differs increasingly from the self-consistent Green 
function at large separations, making the variational property of 
$E_{\rm{LW}}$ less reliable. We also plotted benchmark 
configuration-interaction (CI) results and the binding curve 
obtained from the self-consistent Green function within the second-order 
self-energy approximation \cite{dahlen05b}, which we were able to 
calculate up to $R=6$.
The second-order results are closer to the exact results than
the $GW$ curve around the equilibrium distance. This was to be expected, 
since the main feature of $GW$A is to 
screen the long range interactions. 
For atoms or small molecules it is more important to take both direct and 
exchange diagrams into account to the same order. Also for the atoms, the
SC-$GW$ results are not particularly close to the CI results, as seen in 
Table \ref{tab:totale}. It should be noticed, however, that the shapes of 
the $GW$ and the second-order curves
are similar to each other and to the CI curve around the equilibrium
bond distance.  We finally note that, like the HF method, self-consistent 
$GW$ is not a size-consistent method, i.e.
the total energy calculated at large separations will not converge to the sum 
of the total energy of the fragments. This is not surprising, since the $GW$A
is similar to HF in that the bare interaction in the exchange self-energy
is replaced by a screened interaction and this screening is not sufficient to
alleviate the deficiency of HF. This is an obvious problem when 
calculating molecular binding energies, and has been discussed in more detail 
in Ref.~\cite{dahlen06}. 

Let us now turn to calculations of atomic energy differences.
It is evident from the shape of the binding curves around the equilibrium
separation in Fig.~\ref{fig:bindingh2}, that SC-$GW$ can produce accurate total
energy differences. Calculations on atoms using the LW
functional have also shown that two-electron removal energies, $\Delta E=
E_{N-2}-E_N$, can be very
accurately given within the $GW$ approximation \cite{dahlen04b}. We
therefore calculated the SC-$GW$ removal energies of Be and Mg, as shown in 
Table \ref{tab:subtract}. We find excellent agreement with the experimental 
results for both \chem{Mg} and \chem{Be}, the deviation being ten times smaller than those
from the HF calculations. This improvement is in keeping with the results 
obtained by Shirley and Martin for $G_0W_0$ calculations on atoms 
\cite{shirley93}.
\begingroup
\begin{table}
\caption {Two-electron removal energies $E_{N-2}-E_{N}$
(in eV) calculated from SC-$GW$, compared to HF values
and the experimental values.}
\label{tab:subtract}
\begin{largetabular}{lcccc}
\hline
System & SC-$GW $ &HF & Expt.$^1$ \\ 
\hline
\chem{Mg} -  \chem{Mg^{2+}} & 22.59 & 21.33 & 22.68 \\
\chem{Be} - \chem{Be^{2+}} & 27.59 & 26.17 & 27.53 \\
\hline
\end{largetabular}
$^1$From Ref. \cite {LLK}.
\end{table}
\endgroup
In Table \ref{tab:ioniz}, we show the ionization potentials obtained from the 
EKT, both from the SC-$GW$ and the non-selfconsistent $G_0W_0$
Green function.
The latter is obtained by iterating the Dyson equation once, starting from
an LDA or HF Green function. 
\begin{table}
\caption{ Ionization potentials (eV) calculated from the EKT, using the 
self-consistent Green function and the Green function calculated from one
iteration of the Dyson equation, starting from $G_{\text{LDA}}$ and 
$G_{\text{HF}}$.}
\label{tab:ioniz}
\begin{largetabular}{lcccc}
\hline
System & $G_0W_0$ (LDA) & $G_0W_0$ (HF) & $GW$ & Expt.$^1$\\  
\hline
\chem{He}        &   23.65    & 24.75 & 24.56 & 24.59 \\
\chem{Be}        &   8.88$^2$ & 9.19 & 8.66$^2$ & 9.32 \\
\chem{Ne}        &  21.06 & 21.91 & 21.77  & 21.56 \\
\chem{Mg}        &  7.52 & 7.69 & 7.28  &  7.65  \\
\chem{H_2}  	 & 15.92 & 16.52 & 16.22 & 15.43 \\
\chem{LiH}       & 6.87 & 8.19 & 7.85 & 7.9 \\
\hline
\end{largetabular}
$^1$From Ref. \cite{LLK} \\
$^2$To be compared with the $G_0W_0$ value 9.25  and the SC-$GW$ value 8.47, reported in in Ref.~\cite{delaney04}
\end{table}
For most of the systems, the SC-$GW$ ionization potentials are in good
agreement with the experimental values, and in several cases 
better than those of $G_0W_0$. This is in contrast
to the results for the electron gas, where 
self-consistency worsens the spectral properties \cite{holm98}.

The results for beryllium differ from those recently published by Delaney
{\it et al.} \cite{delaney04}. We find a smaller difference between
the SC-$GW$ and the $G_0W_0(\text{LDA})$ results, and the latter value is also 
further away from the exact value than reported in Ref.~\cite{delaney04}.
One explanation for this deviation may be that while we obtained the ionization 
potentials from the EKT, Delaney \textit{et al.} calculated them from the 
poles of the Fourier transformed function $G(\omega)$. For the self-consistent
ionization potentials, these methods should give the same result (they do in
fact only differ with 0.2 eV), but for the
$G_0W_0$ Green function it is not obvious that the results should agree.
Another difference is that we have carried out our calculations in a basis of
Slater functions, while the orbitals in Ref.~\cite{delaney04} are represented 
on a grid. The Slater basis was systematically extended until reaching 
convergence with respect to the total energy. We include HF orbitals with very 
large eigenenergies, e.g., for Be states up to 843 Hartree, while for Ne the 
highest orbital energy was 976 Hartree. We found good agreement between 
second-order M{\o}ller-Plesset calculations with our basis sets and highly 
converged results from the literature \cite{termath90}. This does not imply 
simultaneous convergence of other properties such as the ionization potential. 
In Table \ref{tab:conv},
we illustrate the convergence of the beryllium atom for two different
basis sets. The main difference between the sets is that basis I contains
Slater functions optimized for HF calculations \cite{clementi}.
\begin{table}
\caption{Convergence of the beryllium ionization potential, IP, (in eV) and
total energy (in Hartrees) for two different basis sets. The value of 
$l_{\text{max}}$ indicates the maximum angular momentum quantum number used in
the basis.} \label{tab:conv}
\begin{largetabular}{lcccccc}
\hline
 & $l_{\text{max}}=2$ & $l_{\text{max}}=3$ & $l_{\text{max}}=4$ 
&$l_{\text{max}}=5$ & $l_{\text{max}}=6$  & $l_{\text{max}}=7$ \\
\hline
IP: Basis I  & 8.552 & 8.602 & 8.625 & 8.636 & 8.641 & 8.644 \\
IP: Basis II & 8.439 & 8.615 & 8.637 & 8.649 & 8.654 & 8.656 \\
E: Basis I  & -14.6954 & -14.6999 & -14.7016 & -14.7024 & -14.7028 &
-14.7028 \\
E: Basis II & -14.6807 & -14.6998 & -14.7015 & -14.7024 & -14.7027 & 
-14.7028\\
\hline
\end{largetabular}
\end{table}
The uncertainty of $\sim 0.02$ eV in the ionization potential indicated in
Table \ref{tab:conv} is typical for the calculations on atoms presented in 
Table \ref{tab:ioniz}.

\section{Conclusions} In summary, we have solved the Dyson equation within 
$GW$A to self-consistency for a number of atoms and diatomic molecules. We 
have shown that SC-$GW$ gives good total energy differences and ionization 
potentials, significantly improving the HF results. We demonstrated that 
self-consistency improves the $G_0W_0$ ionization potentials for most systems 
studied and has the additional advantage of providing unambiguous results.
Moreover, we have shown that the LW functional gives total energies in 
excellent agreement with the SC-$GW$ energies, at a fraction of the computing 
time. This demonstrates the considerable usefulness of the LW functional for 
estimating the accuracy of various self-energy approximations. 

\acknowledgments
We would like to thank Ulf von Barth for useful discussions.


\begin{thebibliography}{0}

\bibitem{hedin65}
\Name{Hedin L.}
\REVIEW{Phys. Rev.}{139}{1965}{A796}

\bibitem{aryasetiawan98}
\Name{Aryasetiawan F., \and Gunnarsson O.}
\REVIEW{Rep. Prog. Phys.}{61}{1998}{237}

\bibitem{godby98}
\Name{Godby R. W., \and Schl{\"u}ter M.}
\REVIEW{Phys. Rev. B}{37}{1998}{10159}

\bibitem{shirley93}
\Name{Shirley E. L., \and Martin R. M.}
\REVIEW{Phys. Rev. B}{47}{1993}{15404}

\bibitem{ku02}
\Name{Ku W., \and Eguiluz A. G.}
\REVIEW{Phys. Rev. Lett.}{89}{2002}{126401}

\bibitem{delaney04}
\Name{Delaney K., Garc\'{i}a-Gonz\'{a}lez P., Rubio A., Rinke P. \and Godby R.}
\REVIEW{Phys. Rev. Lett.}{93}{2004}{249701}

\bibitem{baym62}
\Name{Baym G.}
\REVIEW{Phys. Rev.}{93}{1962}{1391}

\bibitem{holm98}
\Name{Holm B., \and von Barth U.}
\REVIEW{Phys. Rev. B}{57}{1998}{2108}

\bibitem{schindlmayr98}
\Name{Schindlmayr A., Pollehn T. J., \and Godby R.~W.}
\REVIEW{Phys. Rev. B}{58}{1998}{12684}

\bibitem{lw60}
\Name{Luttinger J. M., \and Ward J. C.}
\REVIEW{Phys. Rev.}{118}{1960}{1417}

\bibitem{dahlen04b}
\Name{Dahlen N. E., \and von Barth U.}
\REVIEW{Phys. Rev. B.}{69}{2004}{195102}

\bibitem{almbladh99}
\Name{Almbladh C. -O., von Barth U. \and van Leeuwen R.}
\REVIEW{Int. J. Mod. Phys. B}{13}{1999}{535}

\bibitem{dahlen05b}
\Name{Dahlen N. E., \and van Leeuwen R.}
\REVIEW{J. Chem. Phys.}{122}{2005}{164102}

\bibitem{basisset}
\Book{Details on the basis sets are available upon request.}

\bibitem{smith75}
\Name{Smith D. W., \and Day O. W.}
\REVIEW{J. Chem. Phys.}{62}{1975}{113}

\bibitem{katrieldavidson}
\Name{Katriel J., \and Davidson E. R.}
\REVIEW{Proc. Natl. Acad. Sci., USA}{77}{1980}{4403}

\bibitem{chakravorty93}
\Name{Chakravorty S. J.,  Gwaltney S. R., Davidson E. R., Parpia F. A. \and Fischer C. F.}
\REVIEW{Phys. Rev. A}{47}{1993}{3649}

\bibitem{vanleeuwen94}
\Name{van Leeuwen R.}
\Book{Ph.D. thesis} \Publ{Vrije Universiteit, Amsterdam}\Year{1994}

\bibitem{li03}
\Name{Li X., \and Paldus J.}
\REVIEW{J. Chem. Phys.}{118}{2003}{2470}

\bibitem{gm58}
\Name{Galitskii V. M., \and Migdal A. B.}
\REVIEW{Sov. Phys. JETP}{7}{1958}{96}

\bibitem{dahlen06}
\Name{Dahlen N. E., van Leeuwen R. \and von Barth U.}
\REVIEW{Phys. Rev. A}{73}{2006}{012511}

\bibitem{LLK}
\Name{Lias S. G.,  Levin R. D., \and Kafafi S. A.}
\Book{Ion Energetics Data in NIST Chemistry WebBook, NIST Standard Reference Database Number} \Vol {69} 
  \Editor {Linstrom P.J. \and W.G. Mallard,} \Year{2003} 
  \Publ{National Institute of Standards and Technology, Gaithersburg MD, 20899 (http://webbook.nist.gov)}.

\bibitem{clementi}
\Name{Clementi E.}
\REVIEW{IMB J. Res. Develop. Suppl.}{9}{1965}{2}

\bibitem{termath90}
\Name{Termath V.,  Klopper W. \and Kutzelnigg W.}
\REVIEW{J. Chem. Phys.}{94}{1990}{96}

\end{thebibliography}

\end{document}